# Machining of Spherical Component Fabricated by Selected Laser Melting: Strategies and Equipment

Amir Mahyar Khorasani
School of Engineering, Faculty of Science Engineering and Built Environment, Deakin University, Waurn Ponds, Victoria, Australia
a.khorasani@deakin.edu.au

**Abstract**

The machining process is the most common method for metal cutting, and especially in the finishing of machined parts. In modern industry the goal of production is to manufacture products at a low cost, with high quality in the shortest time. In this research different biomaterials, machinability properties, surface characteristics, cutting tools, cutting fluids and machining conditions for biomaterials with machinability capability are reviewed. In the first step prosthetic acetabular (PA) hip is designed and printed by using selective laser melting (SLM) process then current limitations on fabrication are analyzed to optimize production process and obtain samples with higher quality. The feasibility of artificial intelligence (AI) in machining is determined and In order to calculate dimensional deviation the effect of tool path on tool deflection is modelled. The main focus of this research is determining the machining conditions on surface quality and osseointegration, work hardening and force analyzing of PA. Also the effect of heat treatment on machinability and mechanical properties of produced parts is determined.

# Contents

**Glossary of Abbreviations and symbols**
**Chapter 1**

| | |
|---|---|
| AM | Additive manufacturing |
| CAD | Computer aided design |
| CAM | Computer aided manufacturing |
| SLM | Selective laser melting |
| SEM | Scanning electron microscope |
| ANN | Artificial neural networks |

## 1.1 Introduction

Any artificial/prosthetic material that has interaction with biological systems is termed a biomaterial. Due to the high demand for biomaterials it has seen a steady and strong growth over its short history and therefore many companies are investigating these materials. A combination of various fields such as mechanical, material and surface engineering, biology, tissue engineering, chemistry and medical science have been called biomaterial science. (1-8). Biomaterials can be fabricated by different processes in metallic and non-metallic materials and they should be a part of biological/living systems or they replaced by damaged organs. There are big markets for these materials that are radically growing owing to the increase in demand for better healthcare services. Improvements in innovative biomedical systems can produce major breakthroughs in the healthcare industry, and advanced manufacturing technologies can propel such innovations (9-13). These materials have complicated geometries, mechanical properties and are difficult to manufacture using traditional and non-traditional production processes (14).

Biocompatibility, osteoinduction, osteoconduction, osseointegration, low toxicity and allergy rate are significant medical peculiarities of biomaterials. Parallel to the mentioned characteristics fatigue resistance, machinability, elasticity, strength, corrosion resistance, wear resistance and toughness are essential mechanical properties for biomaterials used in medical applications. Generally, metallic materials are evaluated as candidates for orthopedic implants, bone fixators, artificial joints; and have high tensile strength compare to other materials. Biocompatible ceramics are termed bioceramics/ceramic biomaterials and they are extremely used in medical applications such as dental, joint replacement coating, pacemakers, kidney dialysis machines, and respirators and bone implants owing to durability, low wear and inflammatory response (15).

By virtue of improving technology and their unique mechanical and medical properties, application and usage in recent years, research concerning biomaterials has resulted in them being classified into five sub-groups. Each category consists of six major materials such as Ti-Based, Mg-Based, stainless steels, Co-Based and bioceramics (16-31). This review paper gives an insight into the machining and manufacturing of metallic and ceramic biomaterials. Different biomaterials in terms of machining strategies, machinability, surface quality, proper cutting tools and fluids will be discussed in order to decrease the amount of failure in manufacturing process as well as after implantation.

The paper is divided into five parts start with different machining strategies to highlight which method is more efficient, then machinability of these biomaterials and analysis of the various factors that change the machining rate are discussed. Surface roughness, texture and related treatments in machining are discussed to increase the quality of surface in terms of corrosion and wear resistance, lubricity, mechanical and physicochemical properties which lead to increase implant durability. Cutting fluids and tools are introduced to present an overview for researchers to identify the best choice when machining metallic and ceramic biomaterials.

## 1.2 Biomaterials machining strategies

Different machining conditions have been applied for the machining of biomaterials that are associated with cutting parameters and the type of cooling process which is discussed in this section.

### 1.2.1 Wet machining

The effects of wet cuttings on metal machining have widely been examined (32-34). Wet machining is the most popular material removal process and has superior characteristics such as decreasing adhesion, friction, cutting energy, operation temperature, eliminating built up edge, increasing tool life and efficiency, lubrication and washing chips. Therefore, this process increases the efficiency and decreases production time and cost (35-41).

### 1.2.2 Dry cutting

Dry machining is economical due to the elimination cutting fluid and being environmentally friendly. It has been used for some biomaterials (37). The friction and cutting temperature in dry machining are greater than in wet machining and can lead to reduced tool life, surface quality and thermal shocks. However, reducing thermal shocks in some tools caused increasing tool life (42). Heat distribution and using high quality tool materials such as coated tool, cubic boron nitride (CBN), cermets, and polycrystalline diamond are two important techniques that can recompense for the effect of eliminating cutting fluid.

### 1.2.3 Cryogenic machining

Machining at low temperatures near $-150^0$c is called cryogenic machining. In these operations liquid gases such as helium or nitrogen are poured onto the cutting zone in order to absorb the cutting temperature (43, 44). Commercial cryogenic coolants used in the machining process are liquid helium, air, liquid nitrogen ($LN_2$), liquid carbon dioxide ($LCO_2$) and solid carbon dioxide (dry ice). In machining biomaterials these coolant fluids reduce friction and dissipate high heat. Subsequently, the chemical reactions between the workpiece and tool, adhesion and diffusion of the tool decline; thus tool life, thermal conductivity and efficiency increase (45-50). In cryogenic machining, the mechanical properties of the workpiece are changed; for instance, hardness, cutting force and strength increase, but in contrast, elongation and fracture toughness decrease. The reason can be related to low temperature that causes material to become brittle and under cutting conditions they have different behaviors (51, 52). The most popular agent in the cryogenic machining of biomaterials is liquid nitrogen that which lubricates the cutting surface and reduces the friction coefficient so tool life increases rapidly and processing time and expenses decrease (53-57).

### 1.2.4 Minimum quantity lubricant (MQL) machining

This method is used in situations where surface quality is a very prominent factor and wet or dry machining is not practicable. In this method lubrication is more important than cooling so in the machining of titanium and nickel the high temperature produced in MQL is not effective. In MQL boundary lubrication on the contact surfaces decreases friction on the contact surface of the tool and workpiece and evaporation of the coolant fluid illustrates that circulation and performance are weak (58-60). In MQL coolant fluid and air are mixed and pass through the nozzle that can have different designs like inside or outside of cutting tool. Also, Lopez et al. (61) illustrate that MQL reduced the consumption of coolant fluid by 95%. A MQL study on titanium alloys shows that built up edge in this operation is observable and the workpiece material became harder at the very low air temperature, which resulted in increasing cutting force and power in comparison with dry machining (62). In MQL flank face machining, tool life is longer compared to the rake face by virtue of the cutting fluid not covering the tool chip face during the spraying of coolant fluid and consequently the cutting force decreased (59, 63).

### 1.2.5 Free machining

This method forms small chips and increases the machinability by breaking the chips into small pieces thus avoiding entanglement in the machinery (64-68). The free machining of biomaterials with Pb also allows for higher machining rates. In the machining of titanium alloys ductility and impact resistance decline, whereas temperature and tool life are modified (69-71). Further studies on dental titanium biomaterials (72, 73) have shown that free machining Ti has improved the dental prostheses material removal process.

### 1.2.6 High speed machining (HSM)

In HSM, by increasing the spindle speed in the range of 30000 to 100000 RPM, the temperature dramatically increases and leads to reduced workpiece material strength and cutting force. Also, it has a negative effect on the hardness of tools and decreases tool life(74-76). The increasing temperature in HSM has direct relation to chemical reactivity between the workpiece and cutting tool that has dire consequences on the process such as tool adhesion, wear diffusion and increasing built up edge. In the dry machining of biomaterials increasing cutting speed and temperature are critical because in some cases it can soften the workpiece and decrease cutting force, but in some other aspects, such as the machining of titanium and nickel, this can result in chemical reactions and excessive tool wear (77). In binderless CBN HSM of Ti-6Al-4V workpiece material strength decreases by increasing the cutting temperature [138]. Large plastic deformation at the primary cutting zone results in increasing temperatures in the HSM of ductile material and increases tool and workpiece adhesion (61). For compensating these problems cryogenic machining is suggested (49, 78). Investigation of HSM on nitinol shows that flank wear is highly related to cutting conditions. According to Figure 1-1 (A) by increasing cutting speed up to 100m/min tool wear is decreased then increased for higher speeds. The reason is associated with longer contact time between cutting tool and workpiece in low speed and high cutting force at higher speeds ranging 100-500 m/min. Another possible reason can be attributed to the reactivity of the work material with the tool material and accelerate the adhesion wear. Figure 1 (B) illustrates that by increasing feed rate due to increasing cutting forces tool flank wear increased and Figure 1-1

(C) shows that the value of tool flank wear is a function of increasing depth of cut and chip-tool contact area.

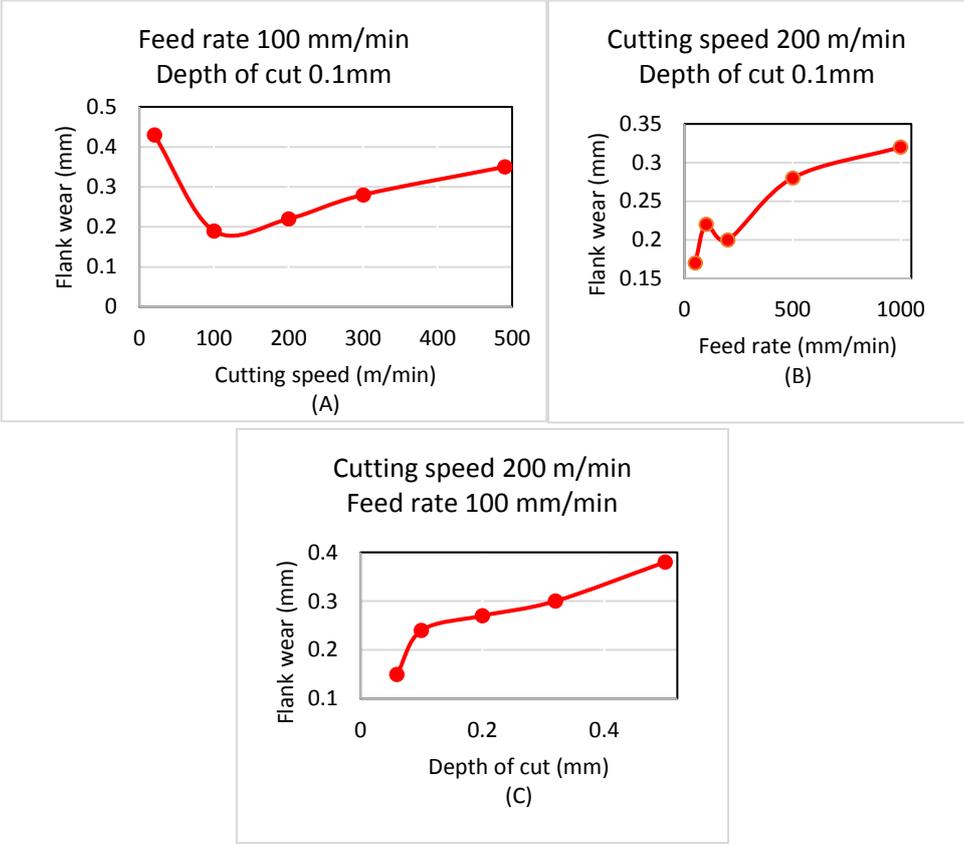

Figure 1 The effect of cutting condition on tool flank wear (79)

### 1.2.7 Air cooling machining

Chilled and compressed air for cooling in machining has been a subject of much research during recent years (60, 80-84). One of the most important outcomes of using air cooling machining of biomaterials is increased tool life and decreased manufacturing time. Also, in air cooling of Ti machining, the cutting force decreases in three dimensions due to decreasing friction. Decreasing cutting forces lead to reducing tool deflection and vibrations to improve surface quality and dimensional deviations (60, 84, 85). Using air cooling machining reduces the amount of adhesion and eliminates built up edge. As a consequence, this method has a superior surface quality than using minimum quantity lubricant and dry machining (60, 62, 86).

### 1.3 Machinability Properties

Machinability identifies which material can be machined with acceptable surface roughness, low surface defects and tool wear; in other words, the rate of material resistance against cutting or material removal is called "machinability". Materials with good machinability require low power to cut, can be cut quickly, are easy to obtain a good finish, and do not wear the tooling much; such materials are said to be free machining.

In the fabrication of implants, traditional material removal methods compare to non-traditional approaches have some superiorities such as a low dimensional deviation, thermal residual stress, high surface quality and efficiency. Moreover, traditional methods using computer numerical control (CNC) controllers have the capability to produce implants with the same geometry as human's organs. Traditional machining of metals, ceramics and polymers is performed to produce ideal surfaces and rough surfaces with controlled textures. Machining processes are important for generating structured surfaces in many biomedical applications (87). Machinability of biomaterials is highly dependent on independent variables that are: tool material, coating, tool shape and geometry, tool sharpness, workpiece material and its processing history, cutting speed, feed rates, depth of cut, cutting fluid, characteristics of the machine tool and the type of workpiece holding (38). In this section, based on these classifications machinability of metal and ceramic biomaterials is discussed.

### 1.3.1 Ti-Based Alloys

Titanium is the most popular biomaterial because of its high tensile strength, hardness, corrosion resistance and in contrast to Cr, Co and Ni it has no side-effects in the human body. "The lightness of titanium (4.5 g/cm$^3$ compared to 7.9 g/cm$^3$ for 316L stainless steel, 8.3 g/cm$^3$ for cast CoCrMo, and 9.2 g/cm$^3$ for wrought CoNiCrMo alloys) and good mechanochemical properties are salient features for implant application" (18, 88). Figure 2 illustrates different Ti-Based biomaterials elasticity module.

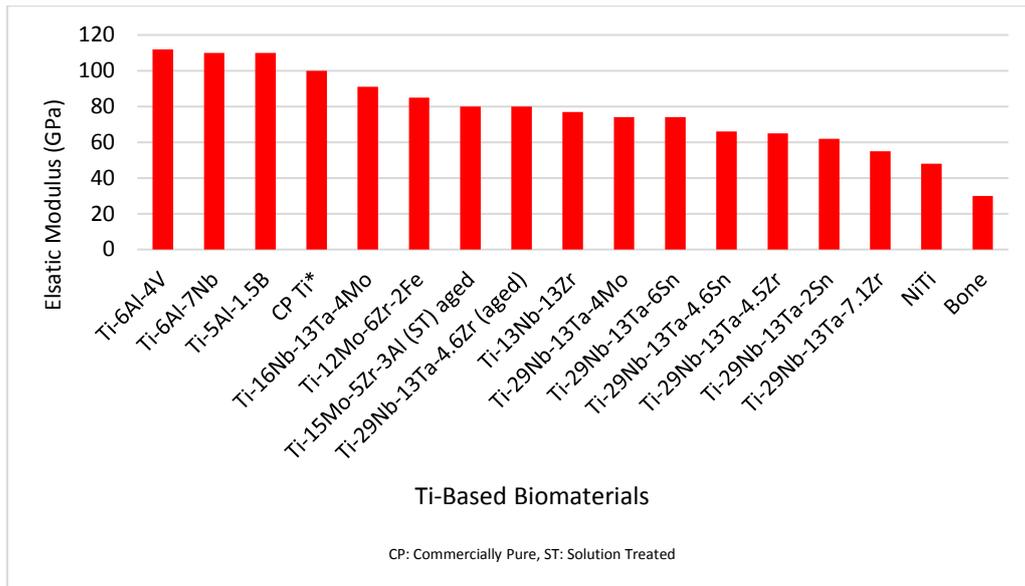

Figure 2 Important Ti-Based Biomaterial elastic module (89)

Progress in the machining of Ti and Ti-Based alloys are not kept due to high temperature chemical reactivity, strength and low thermal conductivity. Approximately 80% of generated heat during machining of Ti-Based alloys are transferred into the tool because low thermal conductivity of Ti prevents any heat transferring from a workpiece. The value of conducted heat into the cutting tool for steel is about 50%. Moreover, extremely steep temperature gradients occur and the heat affected zone is smaller and much closer to the cutting edge owing to thinner chips and make a very thin flow between cutting tool and the chip (about 8μm while this value for iron is about 50μm). Another problem in the machining of Ti-Based alloys is higher cutting pressure that is reported three to four times higher than steels. This is associated with the small contact area for chip and cutting tool in rake face and higher hardness/resistance of Ti to deformation at elevated temperatures. Figure 3 shows cutting pressure for Ti-6Al-4V, steel CK 53 and Nimonic 105.

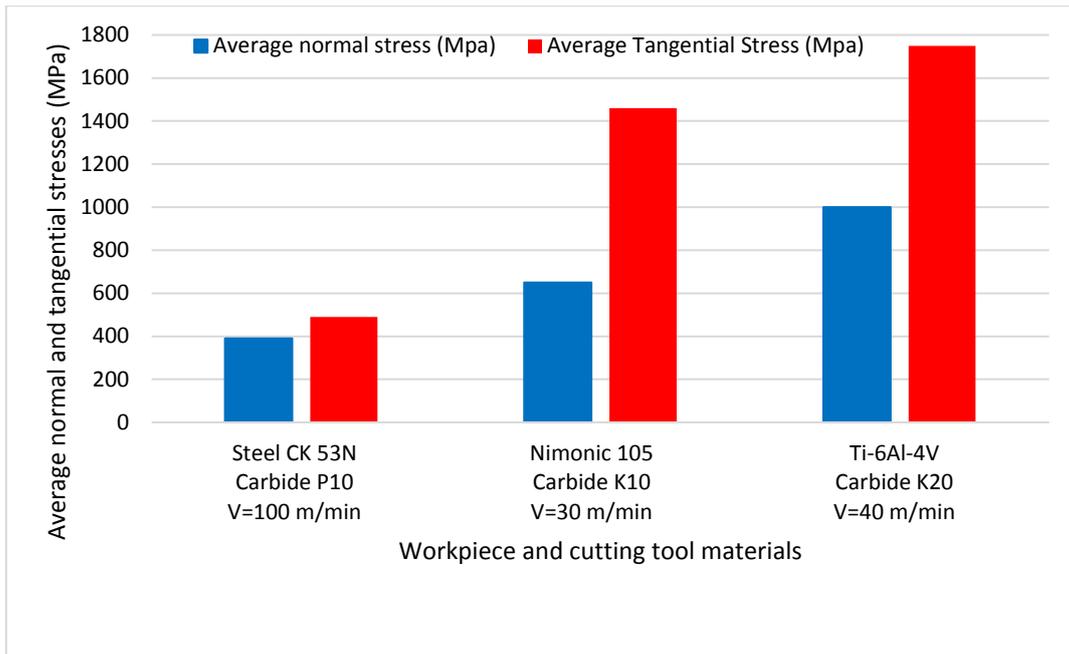

Figure 3 Average cutting stress for different materials and cutting tools [64]

Low elastic modulus of Ti results in increasing the chatter in the machining operation. In a continuous cutting pressure Ti deflect (about two times more than steels) and make promote the spring-back phenomenon and leads to premature flank wear (90).

Ezugwu (91) presents an overview of major advances such as using coated tools and cryogenic cooling in the machining techniques that led to decreasing cutting zone temperature and increases in productivity. Hence, lower manufacturing cost, without adverse effect on the surface finish, surface integrity, circularity and hardness variation of the machined dental material component. This research showed that the use of inert gases due to the poor lubrication and thermal conductivity has negative effect on tool wear and workpiece quality. Moreover, in machining of Ti-Based biomaterials low thermal conductivity is a noticeable issue for CBN and ceramic tools thus using cutting fluid (gas coolant) is recommended.

The integration and novel approach of the tool path generation and simulation for milling in the field of dental technology on Ti alloys was studied by Gaspar and Weichert (92). They optimized the tool path in terms of path length and feed rate thus machinability and efficiency increased. Changing tool path and feed rate resulted in modifying cutting force and decreasing tool wear. The machining of Ti biomaterials and its alloys illustrates that the efficiency of this process when compared to stainless steel and ceramics is dramatically lower, but by using multi-layered coated CBN and tungsten carbide or tungsten cobalt (WC/Co) inserts efficiency increases and wear zones are reduced. Indeed, free machining titanium led to increased tool life and decreased cutting temperature while it had a negative impact on ductility and impact resistance (69-71).

The investigation of 37 patients with 88 titanium restorations (93) has shown that it can be assumed, machined titanium restorations are suitable for clinical use, although not for all problems. Under some machining situations, the machining of titanium alloys produces micro and nanoparticles that are hazardous to the environment and health. The particles are made due to chemical reaction between cutting tool and workpiece in special temperature range. The recommended solution for reducing these contaminants is using HSM. Furthermore; Nb, Ta and Zr are the favourable, non-toxic alloying elements with low rate of particle emission for Ti-Based alloys in biofabrication (94-96). The deformation mechanism in the machining of Ti-Based alloys is complex and abrasion, attrition, diffusion–dissolution, thermal crack and plastic deformation are the main tool wear mechanisms. These problems are related to low thermal conducting and high hardness of Ti-Based biomaterial that led to raising thermal shocks in cutting zones, decreasing yield stress and increasing chemical reaction between cutting tool and workpiece (97).

Machining titanium alloys at high speeds can be beneficial from a particle emissions point of view, but because of their high strength to weight ratio, low thermal conductivity and high interfacial temperature the tool wear rate rapidly increased. For solving the problems and improving high thermal durability, special materials, including hexagonal boron nitride (HBN), ceramic, diamond, and CBN tools are used. In HSM Ti-Based biomaterial by increasing the cutting speed, better surface finish, less work hardening and chip cross-sectional area or chip thickness and a lower cutting force or load was obtained (79, 98, 99). Another way of increasing tool life and decreasing cutting force is to use high-pressure cutting fluid and 5 axis machines respectively that are tested experimentally in the machining of Ti–6Al–4V. High pressure coolant fluid controls machining temperature and 5 axis movement decreases cutting force by making better contact area for cutting tool and workpiece (100, 101). Experimental machining of Ti-Ag and Ti-Cu alloys (102, 103) demonstrates that the cutting force in the machining of titanium is rapidly reduced by adding silver. In Ti-Based alloys milling the average cutting forces are linked to tool geometry and cutting conditions and by optimizing the flute angle to near $20^0$, the force reduces. Decreasing in machining force led to improving surface quality, which includes roughness and cracks, tool life as well as cutting zone temperature (Figure 4 shows the effect of flute angle on cutting stress and force) (104).

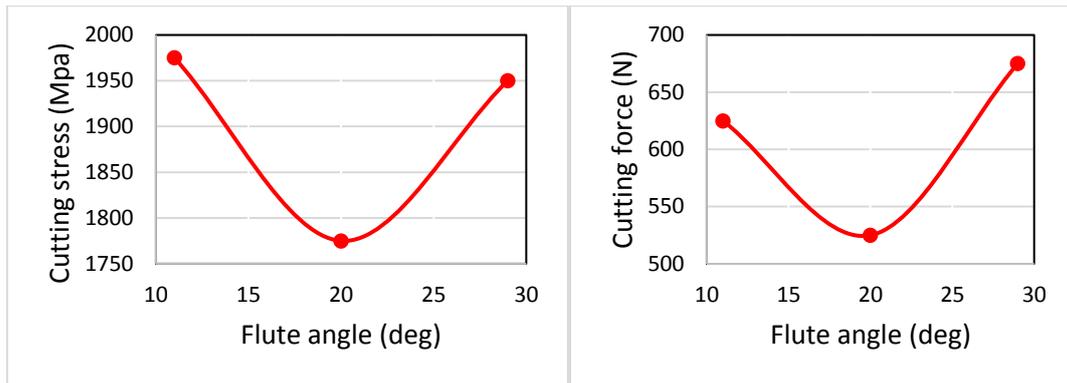

Figure 4 The effect of flute angle on cutting stress and force in the machining of Ti-6Al-4V (104)

Studies on high-speed dental, free machining titanium that have been performed by Tira and Feyerabend et al. (72, 73) illustrate that this operation has superior properties for dental prostheses. In free machining of Ti, the cutting effectiveness of the rotary cutting instruments might be improved, tool life increased and more complicated prosthetic designs might be solved. Indeed, these studies have shown Ti containing lanthanum (due to crystal structure and modifying modulus) appears to be a good alternative for biomedical applications, in machining processes. Studies on the machinability of cast titanium and Ti-6Al-4V (105, 106) illustrate that each material removal process largely depends on the machining conditions when compared to Ti 5553 and Co-Cr alloys, CP-Ti and Ti-6Al-4V appeared to be much easier to cut or machined. The reason is Ti 5553 has extremely low thermal conductivity and Co-Cr alloys are harder than CP-Ti and Ti-6Al-4V.

Grinding is mostly used in the finishing and producing extremely smooth surfaces so this operation is suitable for biofabrication especially in sliding surfaces such as hip and knee joints. Literature on grindability of Ti alloys (107) illustrates that alloying additions lead to increased grindability by reducing fracture toughness, ductility, tensile elongation and the resistance to crack initiation. For instance, adding copper to Ti-6AL-4V causes an increase in grindability and a slight reduction in ductility. In Ti-Cu and Ti-6Al-4V eutectoid and α+β structure increase resistance to plastic deformation and help to enhance grindability by breaking more easily in forming chips indeed, ductility decrease so machinability improve. Generated heat in machining of Ti due to low thermal conductivity remains in the workpiece and tool surface can be improved by adding Cu that has better thermal conductivity. However, adding Cu to Ti increase the hardness and cutting forces therefore, has negative effect in this viewpoint. (108). Ti with 10% Cu alloys have a high rate of grindability and a low rate of wear compared to CP-Ti. Likewise, the grindability and chips ground of Ti–6Al-4V and Ti–Cu alloys with lower Cu concentrations are similar. Another way for increasing the grindability of titanium is by adding 20% Ag, 5% Cu, and 10% Cu alloys so alloys of copper or silver improve microstructure, heat conductivity and subsequently the grindability of Ti, particularly at a high speed (109-111). Ti–40Zr alloy has a great contribution for use as a dental machining alloy and the grinding rate of the Ti–40Zr alloy at 500 m/min was about 1.8 times larger than that of CP-Ti while the grinding rate of Ti–10Zr–5Cr at the speed of 100m/min is about 2.6 times more than for CP-Ti. Adding Zr and Cr balance mechanical properties such as transferring β phase and improve thermal conductivity of CP-Ti. Also, the grinding ratio of these alloys is significantly higher than CP-Ti (112, 113). The alloying elements Nb, Mo, Cr and Fe greatly contributed to increase the grinding ratio under all grinding conditions presumably due to increasing thermal conductivity and the brittle nature of the alloy containing the ω phase in the β matrix. Generally, the grinding rate for all metals is largely dependent on cutting parameters such as grinding speed (114). Ohkubo et al. (115) studied on grindability of biomaterials and they found that the ease of grinding was found in the order of Ti–6Al–4V =Type IV gold > free machining = CP-Ti > Co–Cr. The grindability of the gold alloy was similar to that of Ti–6Al–4V, whereas the Co–Cr alloy because of high hardness, cutting force and wear had the lowest grindability.

Ti-Cr alloys are another subject for bioresearch because of its usage in biomaterials. It is found that Ti–10Cr, Ti-20Cr-X alloy X (X = Nb, Mo, Zr or Fe) Ti–5Cr, Ti–5Cr–0.1Fe, Ti–5Cr–0.5Fe and Ti–5Cr–1Fe have a high grindability rate, especially at 1000 m/min, which is probably due to the brittle nature of the alloy (116-118). Also, "In machining of Ti, the elevated temperature and high strain at tool-workpiece interface may alter workpiece microstructure and result in β to α phase transformation. During phase transformation, some intermediated phase such as ω phase may form which is highly brittle and hard to machine, and it could reduce the fatigue life of machined components" (119).

Hard to machine biomaterials are classified into three groups, ductile biomaterials, non-homogenous biomaterials and hard biomaterials, shown in Figure 5. These individual material groups are discussed in subsequent sections.

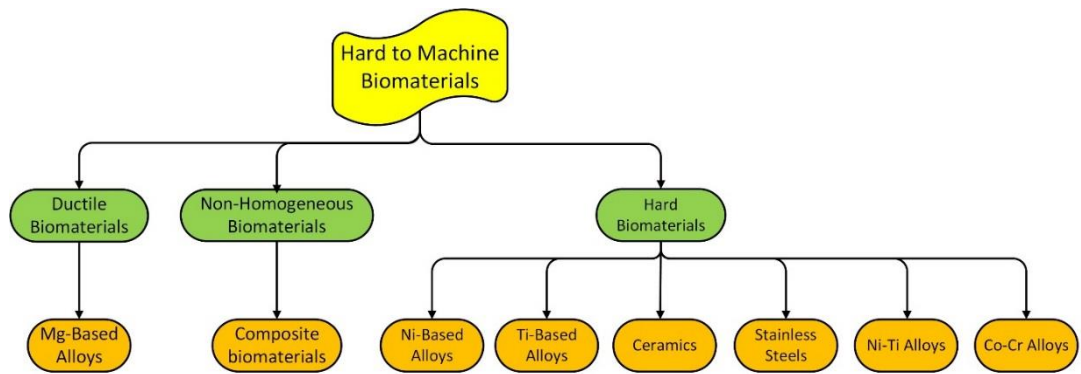
Figure 5 Hard to machine biomaterials (27)

### 1.3.2 Mg-Based Alloys

Magnesium has exclusive properties similar to natural bone, a natural ionic presence and is lightweight. Magnesium alloy has a density of 1.74 g/cm$^3$-that is one of the lightest materials- which has unique properties of high specific strength and high specific stiffness and is utilized in many applications such as biomedical engineering (120-123).

Mg alloys like Mg-Ca0.8 have improved machinability characteristics and obtained chip thickness about 1 micrometer at high speed. Better machinability can be attributed to Ca thermal conductivity that is 8 times more than Ti and prevent thermal shocks in cutting zone (124). Adding 2% Al to AZ series cast Mg alloys led to maximum tensile properties and cutting forces and tool deflection tended to reduce so cutting temperature and the machinability increased (120).

### 1.3.3 Stainless Steels

Steel alloys with a minimum of 10.5% chromium content by mass have been called "stainless steel" and they are a great contribution to the semiconductor and biomedical industries. The most commonly used stainless steel grade, Austenite 316L, has better corrosion resistance against chloride solutions and has significant contributions to medical applications, like bone plates and screws, implants and spinal rods (125).

The effect of the feed rate on chip formation and tool wear of cast stainless steel with hot isostatic pressed (HIPed) NiTi coating was analyzed by Paro et al. (126). In pseudo-elastic material high strength strong deformed layer and work hardening lead to machinability difficulties in cutting process. In addition, In HIP process Fe diffusion on NiTi and enriching Cr decrease tool life and machinability thus effective cutting speeds and feed rates were recommended in order to optimize cutting forces, temperature and tool life without a decrease in coating properties. Figure 6 shows the effect of cutting parameters on surface quality in turning operation of stainless steel. As it can be seen in Figure 6 (A) by decreasing the feed rate and cutting speed the value of roughness is decreased. Decreasing feed rate and depth of cut led to decreasing surface roughness Figures 6 (B), while the interaction of cutting speed and depth of the cut shows (Figure 6 (C)) that the best surface quality is obtained by decreasing the depth of cut and increasing cutting speed.

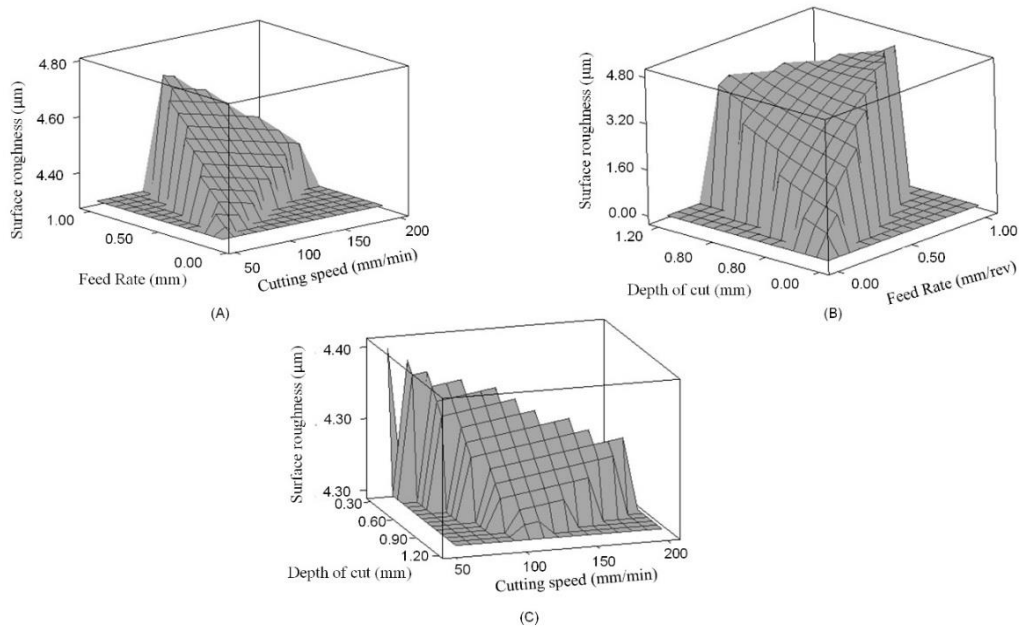

Figure 6 The effect of cutting condition on surface roughness (127)

Modeling of the cutting edge geometry of medical needles that are produced by heat-treatable stainless steels have shown changing inclination angle of the needles resulted in reduction of pain and trauma by the patient. To change this angle using convex grinding wheel and 5 axis grinding is recommended, but using convex tool gives constant inclination angle, while second method is more flexible because cutting contact surface and subsequently cutting force and temperature are controlled (128). Using rapid prototyping and machining of stainless steel illustrates that the machining time has to be improved and tool selection (number and paths) is important to achieve high quality surfaces and machinability. Different tool path leads to changing in force component cutting temperature and tool life (129). Austenite 316L has better machinability compared to that of duplex alloys and the chisel edge area responsible for surface penetration demonstrated a shorter amount of wear on the 316L tools in comparison to duplex; indeed, no flute damage was found in the drilling of austenite316L. The reason is related to gamma phase of iron in FCC structure that is softer and more ductile than duplex alloys so produced force and temperature during machining process are optimized (130).

### 1.3.4 Co-Based Alloys

Co-Cr alloy when used as a biomaterial is required to possess high resistance to wear, corrosion and biocompatibility. "These materials are usually called Co-Cr alloys and divided into two groups: one is the CoCrMo alloy, which is usually used to cast a product, and the other is CoNiCrMo alloy, which is mostly wrought by (hot) forging. The first type has been used in dentistry and joints and the second category has been used for prostheses for heavily loaded joints (such as the knee and hip) (18).

Soft computing techniques have been used by Vera et al. (131) to optimize a dental milling process on Co-Based specimens. They used artificial neural networks ANNs to model surface roughness and production time then optimized it by genetic algorithm and validate it by ANNs. The results of this research led to minor time errors for the manufacturing of dental pieces and minor erosion thus, proved soft computing is appropriate devise for modelling of milling process in dental application. Milling of Co-Cr-Mo dental alloys by using diamond coated tools exhibited increased tool life and decreased friction. Diamond is a good conductor of heat because of low phonon scattering and strong covalent bonding which is five times more than copper. As a result of using diamond coated tools, the machinability of Co-Cr-Mo increased dramatically (132).

Co-Cr grinding using 8000 grit led to producing an average surface roughness of 7 nm and superior surface hardness, biocompatibility and machinability. Figure 7 demonstrates that increasing the number of wheel grits (NWG) in the grinding process resulted in uniform cutting contact areas and force on workpiece and produces better surface quality. These privileges are also related to diffusion of several types of elements in cutting operation and coolant fluid compositions (133).

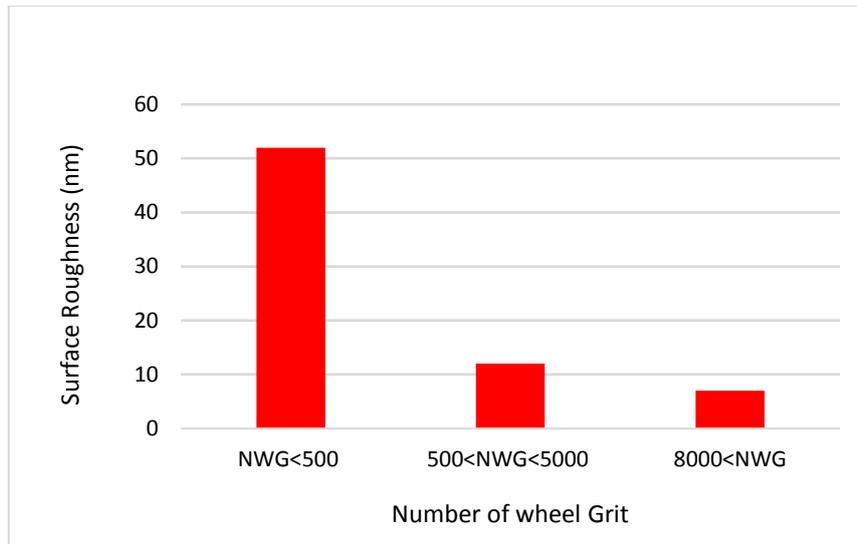

Figure 7 Surface roughness of finished Co-Cr alloy versus wheel grit (133)

### 1.3.5 Ceramics

Ceramic materials are refractory, nonmetallic, polycrystalline compounds, solid and produced by the action of heating and cooling and they have partly and fully crystalline or amorphous, glass like structure, usually inorganic, including silicates, metallic oxides, carbides, and various refractory hydrides, sulfides, and selenides.

Recently, ceramic materials have been given a lot of attention in biofabrication since they have some highly appropriate properties for certain applications. Ceramics have been used in biomedical applications such as dentistry for dental crowns by reason of their inertness to body fluids, high compressive strength, and good aesthetic appearance in their resemblance to natural teeth. Due to their biocompatibility, high specific strength as fibres, they are utilized for reinforcing component for composite implant materials and tensile loading applications such as artificial tendon and ligament replacements (18). Ceramic biomaterial with machining capability is divided into Aluminium Oxides (Alumina), Zirconium Oxides (Zirconia) and Glass-Ceramics which are discussed in this section (134).

The machining of ceramics has problems. For instance, vibrations can lead to a cracks and fracture because of the brittle nature of this material and by reason of low thermal conductivity of them thermal and texture damages are another issue (135). To solve the mentioned problems, the novel direct ceramic machining process by using diamond tools and controlling cutting condition for all-ceramic dental restorations with high mechanical strength is recommended. Indeed, using coolant fluids may lead to thermal shocks on cutting tool so gas cooling system is needed to compensate low thermal conductivity and prevent thermal damages in diamond tool (136). Drilling, milling and turning of alumina/cyanoacrylate green ceramic compacts have shown that the conventional WC tools can be used for this material with a low rate of wear. Green ceramic has suitable characteristics for green machining and no cutting fluid is needed because the compact structure did not show any signs of thermal degradation. By using the finite element method the place of maximum cutting load is determined and by using 5 axis CNC machining cutting forces and tool deflection and damage can be minimized (137-139).

Machining variability impacts on computer aided design and manufacturing (CAD–CAM) ceramics (140, 141) proved that IPS Empress 2 framework ceramic has a better surface quality and machining characteristic than Vita Mark II ceramic that is related to brittle structure and better thermal conductivity of IPS. Investigations on nano-crystalline hydroxyapatite (HA) bio-ceramics showed that these materials have high machinability characteristics that is related to brittle fracture without plastic flow. Observations demonstrate that all cutting parameters have direct effects on surface roughness of this biomaterial, but the feed rate has the greatest impact on cutting force tool deflection and surface quality (5). In the ceramic category glass-ceramics and ceramic green sheets made from viscous polymer processing possess high green strength and good machinability. In viscous polymer processing appropriate solvent and adequate applied pressure are two important factors that influence on the interfacial bonding, strength, resistance to cut and process temperature thus, optimizing these factors lead to improving machinability (142-144). Green machining of ceramics on alumina dental crown due to compact structure, appropriate interfacial bonding and green strength proved that by optimizing cutting parameters cutting force and temperature can be minimized therefore, dental crowns were successfully fabricated with high surface quality and low dimensional deviation (145).

### 1.3.6 Other Biomaterials

This category contains other biomaterials such as composites, dental restoration, nacre, zirconium, and nickel that are used in manufacturing and medical applications.

"In dentistry, amalgam is an alloy of mercury with various metals used for dental fillings. It commonly consists of Hg (50%), Ag (~22-32%), Sn (~14%), Cu (~8%), and other trace metals"(146). Hg has low thermal conductivity (8 W/Mk) and high thermal expansion while Cu and Ag have high thermal conductivity (about 50 times more than Hg) so they improve thermal problems such as shocks, tool wear and surface quality.

 "The NiCr alloys in dentistry are generally used for porcelain veneered and without veneered crowns, fixed and removable partial dentures and bridgework. The corrosion resistance of the NiCr-alloys is provided by the chromium content which produces a passive oxide layer on the surface". These layers increase hardness, tool wear and cutting forces so machinability decreased. Indeed, during dry machining of this alloy Ni carbide appears due to thermal oxidation that is hard component and declined machinability (19).

"NiTi is categorized as a shape memory alloy that found interesting applications in vast areas of engineering from aerospace to biomedical; the latter applications are due to its biocompatibility in addition to its unique properties. This alloy has an unusual property in that after metal is deformed they can "snap-back" to their previous shape following heating. The unique properties such as shape memory, high ductility, severe strain-hardening and pseudoelasticity make NiTi an excellent candidate in many functional designs, while the mentioned characteristics show this alloy is difficult to cut. NiTi has more properties of Ti than Ni so in NiTi machining generated heat is not discharged smoothly, leading thermal shocks and tool wear and decrease machinability (52, 89, 147). Platinum and other new elements in the platinum chemical group like rhodium, iridium and palladium have unique properties and resistance against corrosion, but have poor mechanical properties and are suitable for manufacturing of pacemaker tips. Compare to Ti this metal has low hardness and strength and higher thermal conductivity so lower cutting forces and thermal shocks appear during machining operation which lead to increasing tool life. In addition, high melting points prevent any build-up edge and surface defects (18).

Experiments on micro-end milling of NiTi alloys shown that in up-milling the burrs are thinner compared to down-milling. The reason is related to the higher cutting force on up-milling that leads to dislodge particles from the surface of workpiece and making bigger burrs. Indeed, burrs are thicker on martensitic structure because in this structure carbon atoms do not have time to diffuse out of the crystal structure in large enough and make harder material (148).

NiCu and CoPd alloys are utilized for treating prostate cancers due to their magnetic properties. They are implanted in the tumor and produce heat to kill tumor cells without harming adjacent tissue (18). Ni and Co have high hardness and strength which increase cutting force, but compare to Ti they have high thermal conductivity. Adding Cu and Pd (due to lower hardness and strength) modify mechanical properties and machinability of Ni. Fe–35 wt-%Mn alloy is one of the novel biomaterials that has a high resistance to corrosion, high strength and ability to slowly corrode and is used in stent application. This material is produced by powder metallurgy and has good machinability characteristics. Good machinability is related to the ω phase of iron which is softer and more ductile (149).

Tantalum is mainly used for animal implants and due to the thin oxide layer (formed on their surfaces) that prevents further oxygen penetration, has high rate of biocompatibility. Radioactive tantalum is utilized to treat head and neck tumors (18). High strength and hardness result in higher force and tool wear during machining operation therefore, Tantalum has low machinability and is barely used as alloying element in biomedical applications.

Aluminum is used in bio-applications such as accumulating in soft tissues (150). To produce Ti-6Al-4V and Ti-6Al-7Nb that are the most common alloys in biofabrication, Al, V and Nb should be added to Ti which led to improve thermal conductivity and mechanical properties such as tensile strength. Obviously increasing tensile strength has a direct correlation with cutting force and temperature and decreases machinability. Moreover, oxidation of Al on the surface of the mentioned alloys results in forming stable chemical bonds and increased the hardness.

Gold is used in dentistry due to stability, durability and corrosion resistance. For increasing strength, adding less than 4% platinum is suggested. Softer gold alloys containing more than 83% gold are used for inlays, which are not subjected to too much stress. Harder alloys containing less gold are chosen for crowns and cusps, which are more heavily stressed (18). Au is stable against oxidation, has high thermal conductivity and suitable tensile stress for machining.

One of the novel materials in prosthetic dentistry that is used for the manufacturing of dental parts is zirconium. Finish machining of yttrium cation-doped tetragonal zirconia poly crystals illustrates that machinability and tool wear are greatly related to combined ductile-brittle material removal and adhesion of workpieces respectively. Adhesion prevails tool wear especially when chemical reactions occur between tool and workpiece composition in higher temperature therefore, adhered particles on the surface of cutting tool make dents on the workpiece and decreases surface quality, dimensional deviation and machinability rate (151). Figure 8 shows workpiece material adhesion on the tip of tool in hard turning Ti.

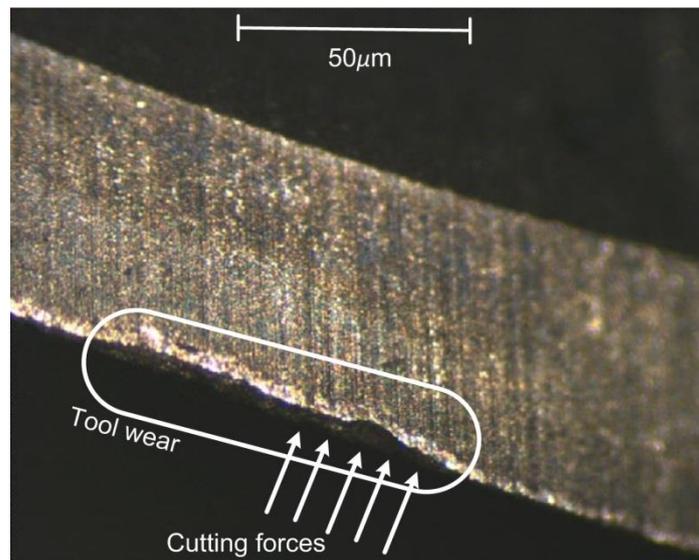

Figure 8 Optical image of solid carbide cutting tool after hard milling of 3D printed Ti (152)

## 1.4 Surface properties in machining

Biomaterial surface has different characteristic such as roughness, patterns, wettability, surface mobility, chemical composition, electrical charge, crystallinity, modulus and heterogeneity to biological reaction. A combination of surface characterization methods is recommended for solving a problem from various perspectives and to provide more comprehensive information about the biomaterials surface (153, 154). In this section the surface roughness of biomaterials is investigated due to its importance in the machining process.

Different surface modification approaches have been developed in all classes of materials to modulate biological responses and improve device performance without altering material bulk properties in various medical and industrial applications (155-161). Surface modification of biomaterial fall into two main groups first one is physicochemical modifications containing alternations to the compounds, molecules and atoms on the surface such as etching, chemical reactions and mechanical finishing polishing or cutting. Second group is surface coating including grafting, non-covalent and covalent coating and thin film deposition (162, 163). Tribological specifications between a soft tissue and textured hard surface are highly associated with the size, density, and orientation of the micro-features. The mentioned characteristics change cell attachment and adhesion so they affect on tribological property (164). Micro-rough surfaces allow early better adhesion of mineral ions or atoms, biomolecules, and cells, form firmer fixation of bone or connective tissue by reason of increasing contact area between implant and living tissues. This phenomenon result in thinner tissue-reaction-layer with inflammatory cells decreased or absent, and prevent microorganism adhesion and plaque accumulation, when compared with the smooth surfaces (165). Moreover, choosing suitable cutting conditions lead to decreasing vibrations in cutting tool and workpiece and subsequently lower surface roughness and defects during machining operations (127, 166-170).

### 1.4.1 Ti-Based Alloys

During machining of titanium alloys, the microstructure of the sub-surface of the bulk material is altered due to plastic deformations and white layer formation. Because of high mechanical and thermal loads on the workpiece in the machining process, work hardening is occurring and the surface and immediate sub-surface of the material becomes harder (171). By Cr-N coating of specimens, surface characteristics and roughness, durability and lubricity of Ti–6Al–4V alloy has been improved. The reason is related to depositing Cr and N atoms on the free gaps of the specimen surfaces and making stronger atomic bonds. In contact area the maximum amount of friction coefficient must be less than 0.15 to decrease friction force and wear. Figure 9 illustrates friction coefficient for different coating and non-coating samples on lubricant test. In finishing and semi finishing dry milling of Ti-Based alloys, using uncoated tool made of carbide (WC–Ti/Ta/Nb–Co) and multilayer chemical vapor deposition (CVD)-coated alloyed carbide (WC–Ti/Ta/Nb–Co+TiN/TiC/TiCN) has the same results on surface quality. This can be related to the delamination of coated layers on high pressure and elevated temperature (172, 173).

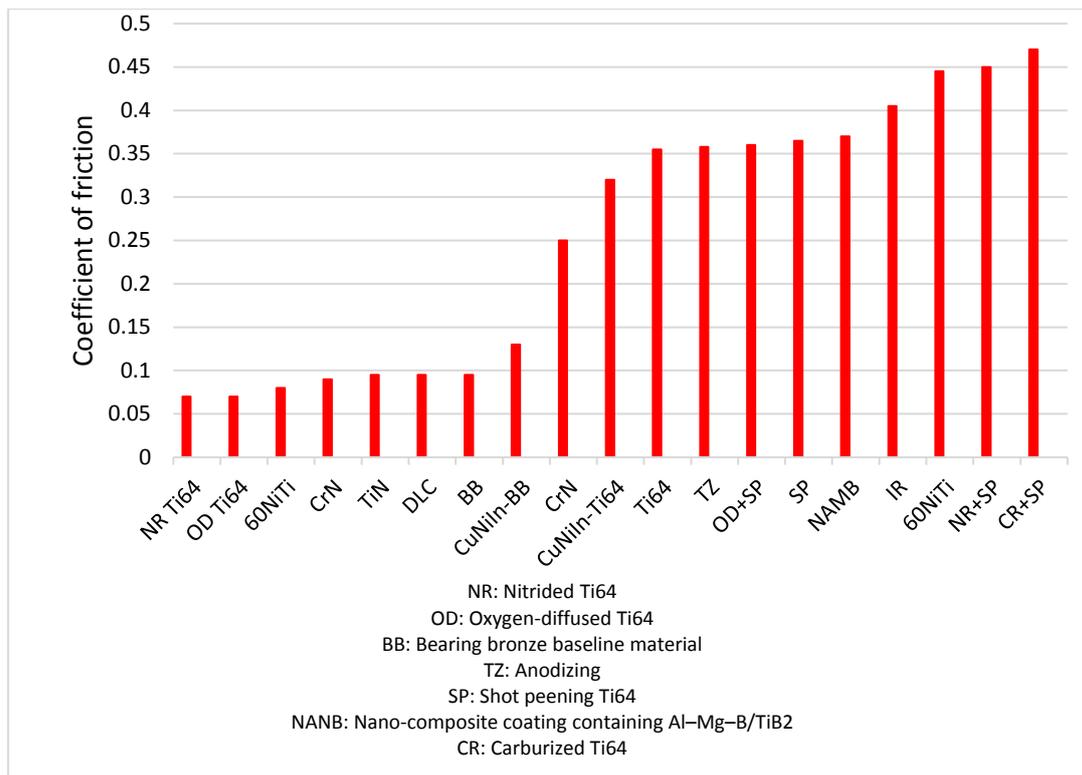

Figure 9 Coefficient of friction for coated and uncoated material under lubricated test conditions (172, 173)

Titanium modification is divided into mechanical, chemical and physical methods according to the formation mechanism of the modified layer on the surface of titanium and its alloys. Nanoscale modification of titanium implants can alter cellular and tissue responses, which is useful for dental implant therapy. A modified surface structure like honeycomb geometry was found to be a promising surface micro pattern. In both nanoscale and honeycomb structures cell attachment improves due to increasing contact surfaces. (174-176). Also, modifying titanium oxide formation due to increasing hardness led to controlling the implants morphology, surface roughness, formation of some compounds that facilitate the wetting and the physical chemistry properties of surface (177). Biomaterial surface roughness impact on human tissue is one of the most important subjects in the machining process. Studies have shown that human bone marrow cells can detect changes in surface roughness of 0.6 microns also implant roughness plays an important role in cells and proteins absorption (178). This research proved that chemical composition and roughness play a predominant role in tissue response. The reason is related to chemical bond that is highly related to chemical composition and cell growth is associated with contact surfaces of implant and cells. In rough surfaces contact area increase, but in very rough surfaces crest and valleys prevent to full contact from cells and therefore cell attachment declined (179, 180). Also, surface roughness and implant chemistry characteristics are significant factors in the bone response to the oxidized implants and mutual force effects between implant and bones. The reason is changing roughness, leads to changing in growth surfaces and subsequently contact forces will be changed. Indeed, chemical bond of bone and implant is highly relied on chemical composition of implant and affects mutual forces (181, 182). Experimentations on zirconia and titanium implant materials illustrate that cell proliferation is occurring around Ti and proved biocompatibility of this metal, indeed roughness, can strongly disturb the relationships between cell proliferation and surface free energy. Surface energy is described the value of disruption of intermolecular bonds so in very rough surfaces cell proliferation reduced. (183, 184). A study on the proliferation of Saos-2 cells and MG-63 cells on finished and rough titanium surfaces has shown that the MG-63 cells, grew more rapidly on rough surface titanium, while the Saos-2 cells grew more on machined titanium than a rough titanium surface. The reason of this phenomenon can be attributed to the shape of these two cells. MG-63 is wider with lenticular shape and has more arms so this cell attaches to rougher implant, while Saos-2 has lamellar shape that cannot attach properly to rough surfaces (185).

Brandao et al. (186) studied the effects of different cooling systems on the surface roughness of titanium Ti-6AL-4V by using two depths of cut, two feed rates, new and worn tools, and three cooling systems as input parameters. Their results proved that although all of the input parameters have a direct impact on the surface roughness, the most important factor is feed rate. Increasing feed rate highly impacts on increasing cutting forces tool deflection and vibration and subsequently surface quality. Also, in the machining of the Ti-Based biomaterial correlation between surface peak parameters ($R_q$, $R_z$, $R_{max}$) and removal torque was detected and further investigations on material removal rate illustrate that acid etching has the best surface roughness followed by the machining, sandblasting and anodizing processes (187) (Table 1). Low cutting forces on acid etching operation produce better surface because during dislodging particles higher forces leading to deterioration of other particle that stick together.

Table 1 mean value ± SD of titanium cylinder surface roughness parameters (187)

| Group | $R_a$ (μm) | $R_q$ (μm) | $R_z$ (μm) |
|---|---|---|---|
| machined | 0.65±0.11 | 0.81±0.17 | 6.09±0.37 |
| acid etched | 0.51±0.10 | 0.71±0.07 | 5.09±0.46 |
| sandblasted | 0.75±0.05 | 0.98±0.04 | 5.55±0.21 |
| anodized | 0.87±0.14 | 1.12±0.18 | 5.14±0.69 |

## 1.4.2 Mg-Based Alloys

The high-speed dry milling on Mg-Based biomaterials has shown surface integrity after machining is characterized by low roughness, highly compressive residual stress and increased microhardness without impressive changing on phase. In Mg-Ca machining rough surfaces showed an irregular degradation with a high degree of resorption therefore, in this operation cutting conditions must be controlled to prevent tool wear, rough surfaces and increasing cell adherence and biocompatibility (188, 189). The machining of Mg-Ca biomaterials proved that the microstructure, including grain size remains stable after burnishing operation. The machined surfaces, due to thermal effects and high cutting force, are harder than the burnished ones down to the deep subsurface around 200 µm because dislocation density and the thickness of near surface deformation layers increased. In the machining of Mg-Ca, subsurface characteristics, such as residual stresses, are highly compressive especially at low burnishing pressure. Also, due to changing dislocation density the subsurface has a strong correlation on corrosion. Machining of Mg-Ca lead to increasing corrosion resistance in terms of increasing dislocation density but using low unsuitable cutting condition result in cracks and decreasing resistance against pitting corrosion (190, 191). Cryogenic machining of Mg-Based biomaterial improved surface finish refinement from 12 µm to 31nm in the featureless surface layer, increased tool life and the surface hardness of the process increased up to 87% compared to the initial values. In this operation friction and thermal stresses decreased so tool life and machining surface improved (192, 193).

## 1.4.3 Stainless Steels

Surface detection of 316L stainless steel samples after polishing treatment demonstrated improved resistance against corrosion. High quality surface properties can be obtained with magnetoelectropolishing, electropolishing and mechanical polishing because of low cutting forces which led to disappearing crack and residual stresses. (194). Surface roughness and wettability of AISI 316L in the milling operation have been investigated by Arifvianto and Suyitno (195). The result revealed that surface mechanical attrition treatment prepared samples surfaces as preprocess and gives smoother surfaces so enhanced surface roughness and a decrease in the droplet contact angle of the AISI 316L surface which is ideal for implant joints.

### 1.4.4 Co-Cr-Based Alloys

On grinding CoCr-Based biomaterial using a #8000 wheel found that the surface roughness of 7 nm is obtainable. Furthermore, in this process superior surface hardness and biocompatibility can be attributed to the quality of surface roughness and using a cutting fluid (133). Enriching Co by tantalum and modifying the surface led to an increase in biocompatibility and lower toxic metal ion release so this alloy is suitable for biofabrication such as prosthetic hip and knee (196).

### 1.4.5 Ceramics

The surface roughness modelling of bio-ceramics in machining process proved that the increasing feed rate led to increasing vibrations and roughness. Also, in the machining of complex ceramic surfaces, the grinding kinematics (cutting conditions) are effective parameter to change surface roughness and cell attachment. Therefore, vibration, feed rate and process kinematic can change biocompatibility by changing surface roughness (197, 198). In the machining of ceramic biomaterials surface roughness, porosity, residual stresses, surface and bulk defects are pertinent parameters to determine the strength and crystalline phases. Optimizing cutting conditions result in enhancing above characteristics thus the quality of produced samples improved (199, 200). A study of the surface roughness in end milling of bone making HA bio-ceramics has shown that the optimization of cutting parameters such as feed rate, cutting speed and depth of cut led to a minimum roughness of 0.64 μm. Feed rate had the highest impact on cutting force, temperature, vibration, tool deflection, surface roughness and subsequently cell growth, attachment and biocompatibility are highly related to this parameter (201). Experiments on the surface characterization of two zirconia based dental materials (Zinelis and Thomas) showed that the surface chemistry, structure and roughness of the implants had significant differences, although both were based on the same composition and received similar surface treatments. "The differences found in the extent of carbon contamination, residual alumina content, tetragonal to monoclinic $ZrO_2$ phase transformation and 3D-roughness parameters may contribute to a substantial differentiation in the cellular and tissue response" (202).

### 1.4.6 Other Materials

The study of the effect of Ni-Cr surface roughness on biocompatibility on living tissues revealed that this dental alloy had the potential to modify the cell density, cell morphology, metabolic activity and cellular toxicity levels by improving surface enhancement operation such as polishing. Also minimum chromium content (16–27%), improved formation of oxide layers to protect surface and therefore increase corrosion resistance (203, 204).

## 1.5 Cutting fluids

Using cutting fluids in the machining of biomaterials results in reduced friction, affecting tools and workpieces, chattering and chip temperature and subsequently preventing welding or adhesion at the contact surfaces, that causes a "built-up edge" on the tool and washing away chips and the reduction of corrosion (35-37, 205). Cutting fluids fall into basic types such as oil-Based, water-Based and gas-Based. Each category contains some different substances that are shown in Figure 10. In the selection of cutting fluids factors like specific manufacturing processes, workpiece and tool materials, processing parameters, compatibility of the fluid with the tool, surface preparation, method of applying the fluid, removal and contaminating of the fluid and costs play an important role (36, 38-41).

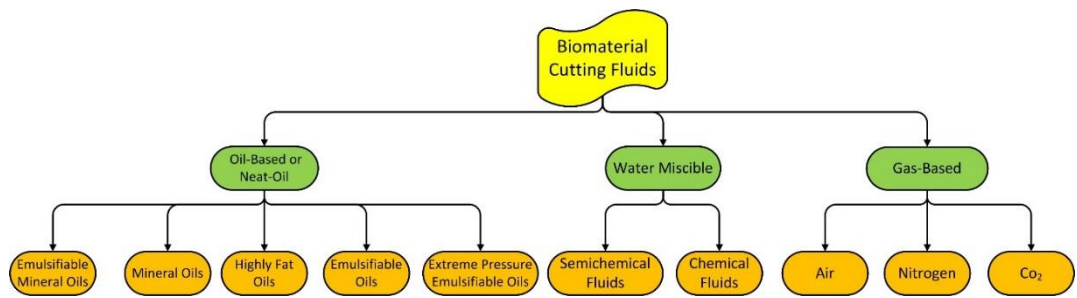

Figure 10 Biomaterial Cutting Fluids (40)

### 1.5.1 Ti-Based Alloys

Titanium has low thermal conductivity (lowest in metals) and is hard to machine so during the machining of titanium cutting fluid plays a prominent role for increasing tool life and surface quality. Extreme pressure, mineral oil, chemical and synthetic coolant fluid have been used for the machining of Ti-Based biomaterial to decrease the cutting temperature and tool wear. Furthermore, in the dry machining of this material localized flank wear is the dominant tool failure and brittle fracture of the cutting edge is observed (38, 206, 207). Indeed, $CO_2$ and liquid nitrogen are used for the machining of titanium biomaterials using diamond tools. These cutting fluids decrease thermal shocks and prevent cracks and fractures on the surface of cutting tools. In the machining of TA48 titanium the best tool is polycrystalline diamond compared with polycrystalline CBN and TiC/TiN/TiCN coated tools. The chemical reaction between carbon and titanium protects the tool against abrasion, decrease diffusion and finally increased tool life by making protective layers (208).

### 1.5.2 Mg-Based Alloys

Gas-Based fluid and light mineral oil of low acid content is normally used for the machining of magnesium biomaterial. Water-Based fluids should not be used on magnesium because of the danger of releasing hydrogen caused by the reaction of the chips with water. Proprietary water-soluble oil emulsions containing inhibitors that reduce the rate of hydrogen generation is the best substance for machining of magnesium biomaterials (206).

### 1.5.3 Stainless Steels

For the machining of stainless steels by turning and milling soluble oil mixed to a consistency of 1 part oil to 5 parts water is recommended. It has been proved that mineral oil-Based and vegetable oil-Based coolant fluid has the same operation level for the machining of steels (38, 209). In machining stainless steels direct liquid nitrogen was employed to avoid heat generation in cutting shear zones (210). Vegetable oil-Based coolant fluid, especially sun flower and canola based, have superior properties over the dry machining of stainless steel biomaterials. Canola cutting fluid with 8% additive has a negative effect on tool life, but has the minimum surface roughness and using sunflower cutting fluid with 8% additive led to increased tool life in the machining of bio stainless steels (211-214).

### 1.5.4 Co-Based Alloys

Due to high specific heat, high thermal conductivity and high vaporization of water miscible oil is recommended for the machining of cobalt and nickel-Based biomaterials (40).

### 1.5.5 Ceramics

Generally for the machining of ceramic biomaterials using coolant fluid is not necessary, but gas-Based coolant fluids are used for machining ceramics. Soluble oil-water emulsions cutting fluid are used during machining process with cemented WC tools (215).

### 1.5.6 Other Biomaterials

For the machining of Cu and Al alloys straight soluble oil and mineral oil are the best cutting fluid that reduce cutting zone temperature and prevent build-up edges (206).

## 1.6 Cutting tools in the machining of biomaterials

In the machining of biomaterials the cutting tool selection plays the dominant role affecting surface quality and mechanical properties of specimens. Thus cutting tools, modifying in design geometries and material and coating with hard materials such as TiC, TiN and TiC-N are the subject of some researches (63, 91, 216-221). Also, using different cutting tools in the machining of the specific workpiece material compensates unsuitable cutting conditions (222).

### 1.6.1 High speed tool steels

The high speed steels are divided into two series containing the T series, which is based on W (with no Mo), and the M series which substitutes Mo. These materials have high resistance to fracture with 67HRC hardness so they are suitable for machining of biomaterials such as Ti and stainless steels through heating to high temperatures (around $1150^0$ to 1250°C). The alloys are composed of 4% to 4.52% Cr, 0.75% to 1.5% C, 10% and 20% W and Mo; they can also have 5% and 12% V and Co respectively which is shown in Table 2. They are strengthened enough for the machining of hard materials like biomaterials (35, 38, 223).

Table 2 sample compositions of some high speed steels (223)

| Grade | Composition (WT. %, balance FE) | | | | | |
|---|---|---|---|---|---|---|
|  | C | Cr | W | Mo | V | Co |
| T1 | 0.75 | 4 | 18 | - | 1 | - |
| M2 | 0.85 | 4 | 6.5 | 5 | 2 | - |
| T6 | 0.8 | 4 | 20.5 | - | 1.5 | 12 |
| T15 | 1.5 | 4.5 | 13 | - | 5 | 5 |
| M42 | 1.05 | 4 | 1.5 | 9.5 | 1 | 8 |

### 1.6.2 Powder metallurgy high speed tool steels

Powder metallurgy high speed tool steel is utilized for milling, turning, reaming, broaching and drilling processes due to its high manufacturing characteristics and increases tool life and refines carbide distribution. Powder metallurgy with at least 5% vanadium content improves wear resistance, non-uniform microstructure, hardness uniformity in heat treatment and tool life so they are highly valued for the machining of bio and hard to machine materials (224, 225).

### 1.6.3 Cast cobalt alloys

Cast cobalt alloys are other materials that are used for the machining of biomaterial. These alloys retain their hardness with machining temperature compared to the carbon and high speed steels. Cast cobalt alloys that are reinforced by chromium and tungsten have a high capacity for HSM. They are now utilized for high speed and high feed rate machining applications by as much as twice the rate possible with high speed steels the composition of this alloy is shown in Table 3. Also, because of their hardness 58-64 HRC these materials are highly resistant against wear and so they can be ideal material for machining of biomaterials. (35, 38, 224).

Table 3 Cast Cobalt alloy composition (224)

| Element | Alloy | |
| --- | --- | --- |
|  | Tantung G, % | Tantung 144, % |
| Cobalt | 42-47 | 40-45 |
| Chromium | 27-32 | 25-30 |
| Tungsten | 14-19 | 16-21 |
| Carbon | 2-4 | 2-4 |
| Tantalum or niobium | 2-7 | 3-8 |
| Manganese | 1-3 | 1-3 |
| Iron | 2-5 | 2-5 |
| Nickel | 7(a) | 7(a) |

### 1.6.4 Cemented carbides

"Cermet and cemented carbide cutting tools consist of hard carbide (or carbo-nitride) grains, bonded or cemented together by up to around 20% by nickel or cobalt, with minor additions of other metals (such as molybdenum or chromium)". Cemented carbides which are used for biomaterial machining are divided in 9 main categories that are TiC (3000HV), VC (2900HV), ZrC (2700HV), HfC (2600HV), WC (2200HV), NbC (2000HV), TaC (1800HV), $Mo_2C$ (1500HV) and $Cr_3C_2$ (1400HV) (223, 224).

### 1.6.5 Cermets

Cermets are a group of powder metallurgy products consisting of ceramic particles bonded with a metal. The metallic component bears ductility and thermal shock, whereas the ceramic component of cermets provides high heat hardness and oxidation resistance. Also, cermet materials that are used for cutting tools are divided into: WC+Co, WC/TiC/TaC+Co, TiC + Ni, TiCN + Ni/Mo, $Al_2O_3$ + TiC.

Titanium carbide/carbonitride is very popular for producing cutting tools and tungsten carbide based materials are named cemented carbides that are superior properties for machining of Ti-Based biomaterial (35, 224).

### 1.6.6 Ceramics

The most modern material for machining processes, especially high speed finishing, and high removal rate machining of difficult to machine material, is ceramic. The combination of boride or carbide and metallic oxide has been called ceramic. This material is well known for its strong resistance against high temperatures and they are produced based on alumina or silicon nitride. Other ceramic cutting tools are zirconia, chromium oxide, magnesia and titanium carbide. Because of their hardness, heat strength and corrosion resistance they are frequently used in the application of biomaterials. (224, 225) Table 4 illustrates ceramic cutting tool hardness.

Table 4 Room temperature hardness of ceramic and WC tool materials (224)

| Tool material | Hardness HRA |
|---|---|
| $AL_2O_3$ | 93-94 |
| $AL_2O_3$-$ZrO_2$ | 93-94 |
| $AL_2O_3$-TiC | 94-95 |
| $AL_2O_3$-SiC | 94-95 |
| $Si_3N_4$ | 92-94 |
| Si-AlON | 93-95 |
| WC-Co alloys | 91-93 |

### 1.6.7 Ultra-hard, super-hard materials

An ultra-hard or super hard tool material has numerous usages in industry and for commercialization. The most common super hard cutting tool materials are diamonds and CBN that are similar in terms of structure and high rates of thermal conductivity. Diamond is a form of carbon and is the hardest substance with 98HRC (35, 38). CBNs are more resistant against air oxidation in normal temperatures. Materials such as copper alloys, carbon fibre composites, tungsten carbide, and some biomaterials like aluminium alloys, polyether ether ketone (PEEKs) and green ceramics have been machined by diamonds. On the other hand, CBN is used for machining other biomaterials like steels, hardfacing alloys, cast iron and surface hardened iron (224). Adding aluminium to CBN tools in the machining of cast iron produces a harder protective layer of aluminium oxide and protects the tool (225). Chemical reactions between workpiece and CBN tools binders result in a decline in tool strength. Using CBN tools in the machining of titanium showed that TiC or cobalt increases operation proficiency. Also, in the machining of Ti-Based alloy using CBN tools, increasing temperature have a direct correlation with chemical reactions between the tool and the workpiece and lead to excessive tool wear (226). Figure 11 shows cutting forces ($F_c$, cutting force, $F_t$, thrust force and $F_z$, feed force) for different coated tools versus uncoated WC cutting tool (71).

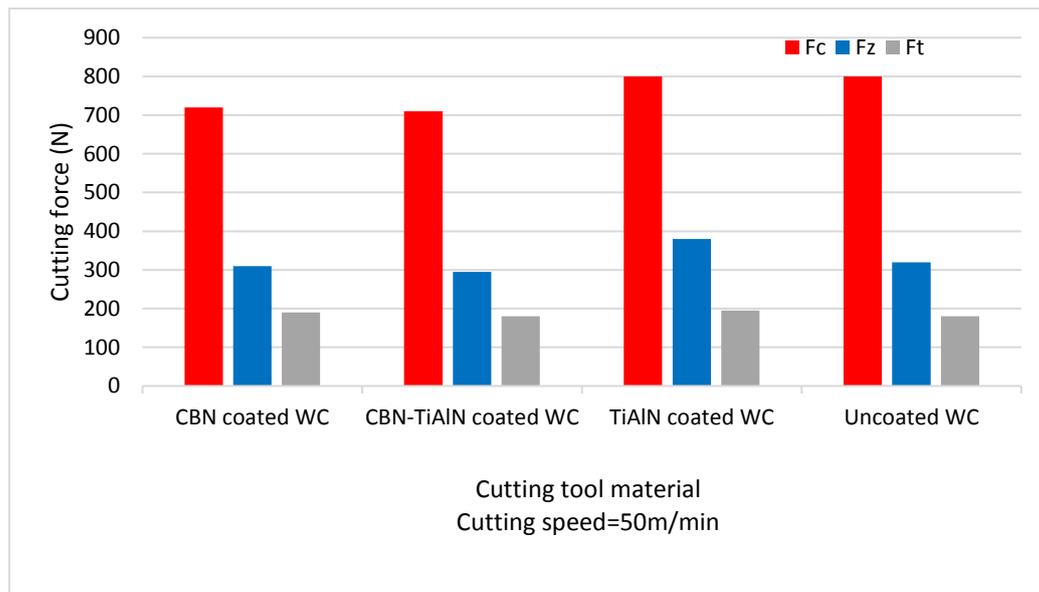

Figure 11 Cutting forces for different coated and uncoated tools (71)

In order to improve the machinability of hard biomaterials using amorphous material such as diamond-like carbon (DLC) coating, polycrystalline cubic boron nitride (PCBN) tools with 90% and 50% CBN ceramic tools respectively; aluminum oxide, titanium carbide and silicon carbide are recommended (227, 228). The net shape manufacturing of green alumina and ceramic via machining showed by using a diamond embedded tool in net shape manufacturing of green alumina (dental crown) the performance and efficiency increased (229). The reason can be attributed to diamond hardness as well as strong covalent bonds and low phonon scattering for diamond that cause good heat conductivity. Figure 12 shows the hardness of cutters in machining of biomaterials.

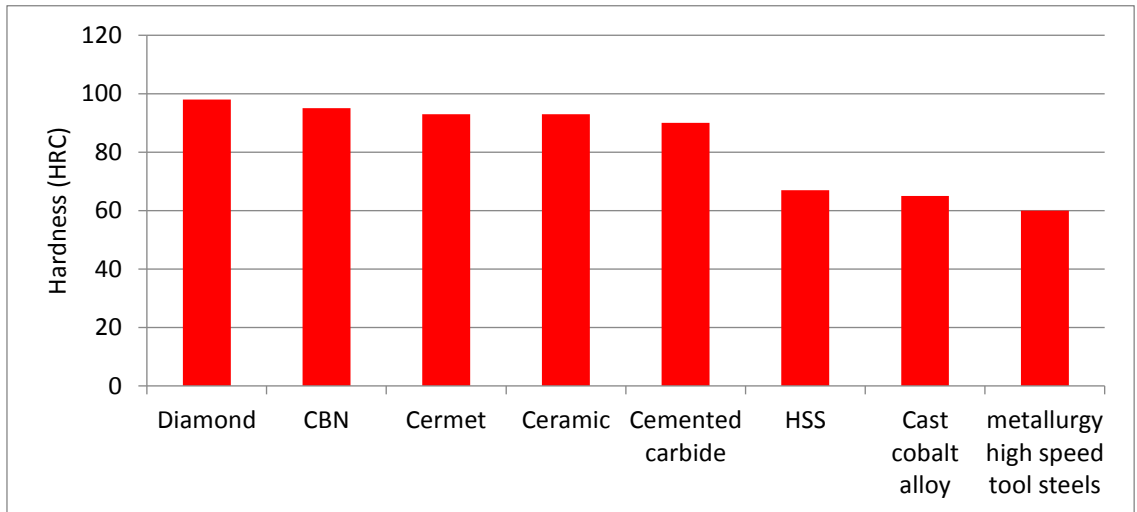

Figure 12 Biomaterial cutting tool hardness (230-232)

## 1.7 Conclusion

In this paper biomaterials, which have machining capability in the traditional machining process, have been reviewed and classified into six sub-groups that are Ti-Based alloys, Mg-Based alloys, stainless steels, Co-Based alloys, ceramics and other biomaterials.

The machining of titanium biomaterials shows the efficiency of this operation due to, high chemical reactivity with cutting tool, low Young's modulus, low thermal conductivity and small chip size is lower than stainless steels and ceramics but it can be compensated by selecting proper cutting tools, fluid and machining strategy. However, CP-Ti and Ti-6Al-4V have better machinability than Co-Cr. Machining of titanium alloys produces some hazardous nanoparticles (from workpiece evaporation) that for decreasing it HSM is suggested, however in this condition increasing temperature to $300^{0}C$ leads to excessive tool wear.

For decreasing cutting temperature and increasing tool life, free titanium machining and high pressure coolant is recommended. Also, the machinability of titanium is improved by adding silver while copper has a negative effect on it, but in grinding Ti-6AL-4V adding Nb, Mo, Cr, Fe, 10% Cu or (20% Ag and 5% Cu) increase grindability. Ti–40Zr, Ti–10Zr–5Cr, Ti–10Cr, Ti-20Cr-X alloy X (X = Nb, Mo, Zr or Fe) Ti–5Cr, Ti–5Cr–0.1Fe, Ti–5Cr–0.5Fe and Ti–5Cr–1Fe have high grind rates compared to CP-Ti.

The machining of Mg-Based alloys biomaterials proved that Mg-Ca0.8 has a better rate of machinability and adding 2% Al to AZ series cast Mg alloys results in increasing machinability.

In machining of stainless steel biomaterials tool life is highly dependent on cutting parameters and surface quality and machinability have a direct relation to tool selection (number and paths). Stainless steels biomaterials are more difficult to machine and need more power than low/middle/high-carbon steels due to their high work hardenability. In addition, the non-free machining steels produce long, stringy chips, which reduce the tool wear and martensitic stainless steels have low machinability due to existing chromium carbides. In this material strength and hardness increased by carbon and nitrogen, which however, cause poorer machinability.

Co-Based alloys biomaterial machining demonstrated that these materials are hard to machine due to low thermal conductivity, a high shear strength and a high work hardening. Using diamond coated tools exhibited increasing tool life, decreasing cutting force and friction in the milling of Co-Cr-Mo dental alloys also high surface quality (7 nm), superior surface hardness, biocompatibility and machinability are achievable in Co-Cr grinding by increasing cutting speed.

The machining of ceramics biomaterial illustrates that texture, thermal damage and vibration in the machining of ceramics cause a decrease in system efficiency. Likewise, green ceramics, IPS Empress 2 framework, sintered alumina compacts and nano-crystalline HA bio-ceramics have superior machinability.

In addition, the machining of zirconia illustrates that tool wear and machinability is related to adhesion of the workpiece. Nacre machining capability, has a dominant characteristic that allows manufacturers to use it in bio-applications. Adding calcium pyrophosphate leads to increasing hardness and improved biocompatibility of Ti-Nb-Mo. Using coated tools decreases friction, increases tool life and operating efficiency.

Future work will be directed toward investigating the machining properties of biomaterials and the relation of the machining conditions on surface quality and biocompatibility. Analyzing the effect of using different machining conditions on dimensional deviations and surface quality are the subject for further research. The modelling of cutting temperature, especially on wet machining needs more attention. Also, there is a lack of research on the machinability of platinum, iron-manganese, gold, Co-Cr and tantalum. Comparing the machinability of biomaterials that are produced from different production processes such as casting, powder metallurgy; 3D printing and so on is another topic that needs further attention.